\documentclass[aps,superscriptaddress,showpacs,floatfix,amsmath,amssymb,nofootinbib,preprintnumbers,
twocolumn]{revtex4-1}

\usepackage{graphicx}

\usepackage{epstopdf}

\usepackage{color}

\newcommand{\be}{\begin{equation}}
\newcommand{\ee}{\end{equation}}
\newcommand{\ba}{\begin{eqnarray}}
\newcommand{\ea}{\end{eqnarray}}

\newcommand{\beq}{\begin{equation}}
\newcommand{\eeq}{\end{equation}}
\newcommand{\beqa}{\begin{eqnarray}}
\newcommand{\eeqa}{\end{eqnarray}}

\newcommand{\bea}{\begin{eqnarray}}
\newcommand{\eea}{\end{eqnarray}}

\begin{document}
\title{Universality of $P-V$ Criticality in Horizon Thermodynamics}

\author{Devin Hansen}
\email{dhansen@perimeterinstitute.ca}
\affiliation{Perimeter Institute, 31 Caroline St. N., Waterloo,
Ontario, N2L 2Y5, Canada}
\affiliation{Department of Physics and Astronomy, University of Waterloo,
Waterloo, Ontario, Canada, N2L 3G1}
\author{David Kubiz\v n\'ak}
\email{dkubiznak@perimeterinstitute.ca}
\affiliation{Perimeter Institute, 31 Caroline St. N., Waterloo,
Ontario, N2L 2Y5, Canada}
\affiliation{Department of Physics and Astronomy, University of Waterloo,
Waterloo, Ontario, Canada, N2L 3G1}
\author{Robert B. Mann}
\email{rbmann@uwaterloo.ca}
\affiliation{Department of Physics and Astronomy, University of Waterloo,
Waterloo, Ontario, Canada, N2L 3G1}

\date{March 17, 2016}  

\begin{abstract}
We study $P-V$ criticality of black holes in Lovelock gravities in the context of  horizon thermodynamics.
The corresponding first law of horizon thermodynamics emerges as one of the Einstein--Lovelock equations and assumes the universal (independent of matter content) form $\delta E=T\delta S-P\delta V$, where $P$ is identified with the total pressure of all matter  in the spacetime
(including a cosmological constant $\Lambda$ if present). We compare
this approach  to recent advances in extended phase space thermodynamics of asymptotically AdS black holes where the `standard' first law of black hole thermodynamics is extended to include a pressure--volume term, where the pressure is entirely due to the (variable) cosmological constant. We show that both approaches are quite different in interpretation. Provided there is sufficient non-linearity in the gravitational sector, we find that horizon thermodynamics admits the same
 interesting  black hole phase behaviour seen in the extended case,  such as a Hawking--Page transition, Van der Waals like behaviour, and the presence of a triple point.
  We also formulate the Smarr formula in horizon thermodynamics  and discuss the interpretation of the quantity $E$ appearing in the horizon first law.
\end{abstract}


\maketitle

\section{Introduction}

 That spacetimes with horizons show a remarkable resemblance to thermodynamic systems has been a subject of study since seminal papers of Bekenstein, Hawking, Bardeen, and Carter 
\cite{Bekenstein:1973ur, Bekenstein:1974ax, Hawking:1974sw, Bardeen:1973gs}. In fact, there is a strong belief that the Einstein field equations, describing the dynamics of gravity,
can be interpreted as a thermodynamic equation of state and have a deep connection with the first law of thermodynamics, e.g.
\cite{Sakharov:1967pk, Jacobson:1995ab, Hayward:1997jp, Padmanabhan:2003gd}. In particular,   
it was explicitly shown that  Einstein equations on the horizon of a  spherically symmetric spacetimes 
can be interpreted as a thermodynamic identity. This was the origin of {\em horizon thermodynamics}  \cite{Padmanabhan:2002sha}.

The original observation for spherically symmetric black holes in Einstein's gravity \cite{Padmanabhan:2002sha} has since been extended to a number of other interesting cases, many of which have been highlighted in recent reviews
\cite{Padmanabhan:2009vy, Padmanabhan:2013xyr}. For example, the horizon thermodynamics has been extended to spherically symmetric black holes
in Lovelock and Quasi-topological gravities \cite{Paranjape:2006ca, Kothawala:2009kc, Tian:2010yb, Sheykhi:2014rka}, $f(R)$ gravity \cite{Akbar:2006mq},
and Horava--Lifshitz gravity \cite{Cai:2009ph},
to time evolving and axisymmetric stationary black hole horizons \cite{Kothawala:2007em, Cai:2008mh}, to horizons in FRW spacetime
\cite{Akbar:2006kj, Cai:2006rs, Gong:2007md}
and braneworld scenarios \cite{Cai:2006pa, Sheykhi:2007gi}. 
More recently the general thermodynamic properties of null surfaces have been investigated e.g. in \cite{Chakraborty:2015hna}.

 In our paper we  concentrate on horizon thermodynamics of black holes.
The basic idea is as follows.
Consider a spherically symmetric black hole spacetime, written in standard coordinates,  and identify the total pressure $P$ with the $T^r{}_r$ component of the energy--momentum tensor of all the matter fields, including the cosmological constant,  if present. The Einstein equations on the black hole horizon can then be regarded as an {\em Horizon Equation of State} (HES)
\be\label{State}
P=P(V,T)\,,
\ee
where $T$ is the temperature of the horizon, identified for example through the Euclidean approach.  By considering an infinitesimal virtual displacement of the horizon, 
one can demonstrate the {\em Horizon First Law} (HFL)
\be
\delta E=T\delta S-P\delta V
\ee
from the radial Einstein equation, where $S$ is the entropy associated with a given black hole horizon and must be independently calculated by other means \cite{Wald:1993nt, Iyer:1994ys}.   The quantities $E$ and $V$ above are respectively interpreted as an energy and a volume associated with the black hole.  We shall consider the nature of these interpretations and their underlying assumptions in what follows.


Interestingly, the idea of pressure and volume as well as that of the equation of state \eqref{State} have in recent years been the subject of much attention in  the {\em extended phase space thermodynamics}
of asymptotically AdS black holes,  see e.g. \cite{Kubiznak:2014zwa, Dolan:2014jva} for recent short reviews. In this framework one identifies  the  cosmological constant as a thermodynamic variable analogous to pressure \cite{CaldarelliEtal:2000, KastorEtal:2009, Dolan:2010, Cvetic:2010jb}.
Its conjugate thermodynamic volume can be obtained via geometric means by generalizing the first law of black hole mechanics in spacetimes that have a cosmological constant \cite{KastorEtal:2009,Dolan:2013ft}.
This in turn implies that the
mass of an AdS black hole is the enthalpy of  spacetime.  This approach  emerged from geometric derivations of the Smarr formula for AdS black holes \cite{KastorEtal:2009} and led to a $reverse$
isoperimetric inequality conjecture  \cite{Cvetic:2010jb}, which
states that for fixed thermodynamic volume, the entropy of an AdS black hole is maximized
for Schwarzchild AdS.  This inequality holds for all known black holes of spherical topology; exceptions exist if this condition is relaxed \cite{Hennigar:2014cfa}.
A very rich and interesting array of thermodynamic behaviour for both AdS and dS black holes then
emerges.  Examples  of the so-called {\em $P-V$ criticality} include a complete analogy between 4-dimensional
Reissner--N\"ordstrom AdS black holes and the Van der Waals liquid-gas system \cite{Kubiznak:2012wp},
the existence of reentrant phase transitions in rotating \cite{Altamirano:2013ane} and Born--Infeld \cite{GunasekaranEtal:2012} black holes, tricritical points in rotating black holes analogous to the triple point of water \cite{Altamirano:2013uqa}, and isolated critical points in Lovelock gravities \cite{Frassino:2014pha, Dolan:2014vba}.
 These phenomena continue to be subject to intensive study in a broad variety of contexts
 e.g. \cite{
Altamirano:2014tva, Kubiznak:2015bya,
Dolan:2012,
BelhajEtal:2012,
SmailagicSpallucci:2012,
HendiVahinidia:2012,
Zou:2014mha,
Wei:2014hba,
Mo:2014wca, Mo:2014mba,
Liu:2014gvf, Johnson:2014yja, Johnson:2014pwa, Karch:2015rpa, Caceres:2015vsa, Hennigar:2015esa, Dolan:2014cja, Maity:2015ida, Zhang:2014uoa, Kastor:2014dra, Wei:2015iwa}.

The goal of this paper is to understand the relationship between these two approaches to gravitational thermodynamics.   Although both  have wider applications, for concreteness  we focus in this paper on spherically symmetric black holes in Lovelock gravity.  After briefly reviewing horizon thermodynamics in this setting \cite{Paranjape:2006ca, Kothawala:2009kc, Tian:2010yb,Ma:2015llh} we
 i)~formulate the horizon equation of state for general $K$-th order Lovelock black holes ii) re-derive the corresponding horizon first law  iii) obtain the corresponding {\it Horizon Smarr Formula} (HSF) and Gibbs free energy and study the associated $P-V$ criticality, and  iv) compare this  procedure and obtained results with the recent advances on extended phase space thermodynamics.    We discuss the  interpretation of the energy $E$ (sometimes referred as horizon internal energy \cite{Paranjape:2006ca}) that appears in both the horizon first law and the HSF we derive, and relate it to the gravitational enthalpy.

Our paper is organized as follows. In the next section we derive the horizon equation of state for a generic Lovelock spherically symmetric black hole.
This equation of state is then `upgraded' to the horizon first law in Sec.~III, where also the associated Gibbs free energy and Smarr relation are studied. $P-V$ criticality is investigated for various Lovelock gravities in Sec.~IV. Sec.~V discusses the relationship to extended phase space thermodynamics. Sec.~VI is devoted to conclusions.

\section{Lovelock gravity and horizon equation of state}

Lovelock gravity \cite{Lovelock:1971yv} is a geometric higher curvature theory of gravity that can be considered as a natural generalization of Einstein's theory to higher dimensions---it is the unique higher-derivative theory that gives rise to second-order field equations for all metric components.  In $d$ spacetime dimensions, the Lagrangian reads
 \begin{equation}
\mathcal{L}=\frac{1}{16\pi G_N}\sum_{k=0}^{K}\alpha_{k}\mathcal{L}^{\left(k\right)} + \mathcal{L}_{m}\,.
\label{eq:Lagrangian}
\end{equation}
Here, $K=\lfloor\frac{d-1}{2}\rfloor$ is
the largest integer less than or equal to $\frac{d-1}{2}$, $\mathcal{L}^{\text{\ensuremath{\left(k\right)}}}$ are the $2k$-dimensional Euler densities, given by
\be
\mathcal{L}^{\left(k\right)}=\frac{1}{2^{k}}\,\delta_{c_{1}d_{1}\ldots c_{k}d_{k}}^{a_{1}b_{1}\ldots a_{k}b_{k}}R_{a_{1}b_{1}}^{\quad c_{1}d_{1}}\ldots R_{a_{k}b_{k}}^{\quad c_{k}d_{k}}\,,
\ee
with the  `generalized Kronecker delta function' $\delta_{c_{1}d_{1}\ldots c_{k}d_{k}}^{a_{1}b_{1}\ldots a_{k}b_{k}}$  totally antisymmetric in both sets of indices, $R_{a_{k}b_{k}}^{\quad c_{k}d_{k}}$ is the Riemann tensor,
and the $\alpha_{\left(k\right)}$ are the  Lovelock coupling constants. In what follows we identify the (negative) cosmological constant $\Lambda=-\alpha_{0}/2$, and set $\alpha_{1} = 1$ to remain consistent with general relativity.
We also assume minimal coupling to the matter, described by the matter Lagrangian $\mathcal{L}_{m}$.
The Lovelock equations of motion  that follow from the variation of \eqref{eq:Lagrangian} are
\begin{equation}
\label{eq:eom}
\sum\limits_{k=0}^{K}\alpha_kG_{\mu\nu}^{(k)} = 8\pi T_{\mu\nu}\,,
\end{equation}
where $G_{\mu\nu}^{(k)}$ are the $k$th-order Einstein--Lovelock tensors \cite{Lovelock:1971yv, Maeda:2011ii}.

 We shall restrict our attention to  spherically symmetric AdS Lovelock black holes, employing the ansatz \cite{Maeda:2011ii}
\be
ds^2=g_{\mu\nu}dx^\mu dx^\nu = \gamma_{ab}(r)dx^adx^b+r^2h_{ij}dx^idx^j\,,
\ee
where the non-trivial part of the metric is described by  a 2-dimensional metric $\gamma_{ab}$ ($a,b=0,1$), while
$h_{ij}$ ($i, j=2, \ldots, d-1$)  stands for the line element of a $\left( d-2 \right)$-dimensional  space of constant curvature $\sigma(d-2)(d-3)$, with  $\sigma=+1,0,-1$ for spherical, flat, and hyperbolic
geometries respectively of finite   volume   $\Sigma_{d-2}$, the latter two cases being compact via identification
\cite{Aminneborg:1996iz,Smith:1997wx,Mann:1997iz}.
The $(a,b)$-components of the $k$th Lovelock--Einstein tensor then are \cite{Maeda:2011ii}
\ba\label{Gabk}
G_{ab}^{(k)} &=&
\frac{k(d\!-\!2)!}{(d-2k-1)!}\frac{(D^2r)\gamma_{ab}-D_aD_br}{r}\left(\frac{\sigma\!-\!(Dr)^2}{r^2}\right)^{\!k\!-\!1}
\nonumber
\\
&&- {\frac{(d-2)!(d-2k-1)}{2(d-2k-1)!}}\gamma_{ab}\left(\frac{{\sigma}-(Dr)^2}{r^2}\right)^k
\ea
where $(Dr)^2=\gamma^{ab}(D_a r)(D_b r)$ and $D^2r = D^aD_a r$.  The remaining $(i,j)$ components can be found in~\cite{Maeda:2011ii}. As long as at least one $\alpha_{k} \neq 0$ for $k> 1$ all possible values of $\sigma$ yield solutions, even if $\Lambda \propto \alpha_{0} =0$.

Consider a black hole for which
\be\label{form}
\gamma=\gamma_{ab}(r)dx^adx^b=-f(r)dt^2+\frac{dr^2}{g(r)}\,,
\ee
with the outer black hole horizon located at $r=r_+$, determined from $f(r_+)=0$.
Employing \eqref{form}, we have
\ba
D^2r&=&\frac{1}{2}\frac{(fg)'}{f}\,,\quad (Dr)^2=g\,,\nonumber\\
D^tD_tr&=&\frac{1}{2}\frac{gf'}{f}\,,\quad
D^rD_r r=\frac{1}{2}g'\,.
\ea
The Einstein--Lovelock equations \eqref{Gabk} then read
\ba
8\pi T^t{}_t&=&\frac{g'}{2r}\sum_{k=1}^K\alpha_k\frac{k(d-2)!}{(d-2k-1)!}\left(\frac{\sigma\!-\!g}{r^2}\right)^{k-1}
\nonumber\\
&&\!\!\!-\sum_{k=0}^K\alpha_k{\frac{(d-2)!(d\!-\!2k\!-\!1)}{2(d-2k-1)!}}\left(\frac{{\sigma}\!-\!g}{r^2}\right)^k\!\!, \label{Ttt}\\
8\pi T^r{}_r&=&\frac{f'g}{2rf}\sum_{k=1}^K\alpha_k\frac{k(d-2)!}{(d-2k-1)!}\left(\frac{\sigma\!-\!g}{r^2}\right)^{k-1}
\nonumber\\
&&\!\!\!-\sum_{k=0}^K\alpha_k{\frac{(d-2)!(d\!-\!2k\!-\!1)}{2(d-2k-1)!}}\left(\frac{{\sigma}\!-\!g}{r^2}\right)^k\!\!.\label{Trr}\quad
\ea

Identifying temperature with surface gravity yields
\be\label{Tdef}
 T=\frac{\kappa}{2\pi} = 
 \frac{\sqrt{f'(r_+)g'(r_+)}}{4\pi}=\frac{f^\prime(r_+)}{4\pi}\,.
\ee
where
\be\label{regularity}
g(r_+)=0\,,\quad f'(r_+)=g'(r_+)
\ee
is required in order that the surface $r=r_+$ be a regular horizon null surface and not a singularity. Noted previously for the Einstein equations \cite{Padmanabhan:2002sha, Kothawala:2007em}, this condition must hold for any Lovelock theory in the spherically symmetric case, and follows from evaluating the trace $ g^{\mu\nu} G_{\mu\nu}^{(k)}$ of each term in the equations of motion \eqref{eq:eom}.   It is straightforward to show that each one of these traces contains a term proportional to the Ricci scalar $R^{(\gamma)}$ for the metric \eqref{form}, which is singular at the horizon unless \eqref{regularity} holds.

Horizon thermodynamics is based on the proposal that the energy--momentum tensor on the horizon is
interpreted as
\be\label{Pm}
P_m\equiv T^r{}_r|_{r=r_+}\,.
\ee
with the assumption that
\be\label{Vol}
V=\frac{\Sigma_{d-2}r_+^{d-1}}{d-1}
\ee
is the conjugate black hole volume.  Note that
$ T^t{}_t|_{r=r_+}=T^r{}_r|_{r=r_+}$ due to the regularity condition \eqref{regularity}.
On the horizon, equation \eqref{Trr} (or equivalently \eqref{Ttt}) thus reduces to
\ba
8\pi P_m&=&
\frac{2\pi T}{r_+}\sum_{k=1}^K\alpha_k\frac{k(d-2)!}{(d-2k-1)!}\left(\frac{\sigma}{r_+^2}\right)^{k-1}
\nonumber
\\
&&-\sum_{k=0}^K\alpha_k {\frac{(d-2)!(d-2k-1)}{2(d-2k-1)!}}\left(\frac{{\sigma}}{r_+^2}\right)^k\,,\quad\ \
\label{horP}
\ea
upon using the regularity conditions \eqref{regularity} and \eqref{Pm} and the definition \eqref{Tdef}
of temperature $T$.

Let us further identify
\be\label{PL}
P_\Lambda=-\frac{\Lambda}{8\pi}=\frac{\alpha_0}{16\pi}
\ee
as the pressure associated with the the cosmological constant, and
\be\label{totP}
P=P_m+P_\Lambda
\ee
as the {\em total pressure} of all the matter fields.
Note that such $P$ is determined from the matter content and is not necessarily positive.
We therefore arrive at
\ba\label{EState}
P&=& {\sum_{k=1}^K\frac{\alpha_k}{4r_+}\frac{(d-2)!}{(d-2k-1)!}\left(\frac{\sigma}{r_+^2}\right)^{k-1}}\nonumber\\
&&\qquad  {\times \left[kT-\frac{\sigma(d-2k-1)}{4\pi r_+}\right] }
\ea
which,  together with the identification \eqref{Vol}, gives the HES for Lovelock gravity, $P=P(V,T)$.
Note that to write down this equation of state one does not need to know the explicit form of $f$.
Furthermore,  equation \eqref{Vol} is an ansatz in this approach that has to be justified (similar to the prescription for temperature $T$) by some other means, e.g. \cite{Parikh:2005qs, Cvetic:2010jb, Ballik:2010rx, Ballik:2013uia}.

\section{Horizon first law \& Gibbs free energy}
\label{Sec:gibbs}

 To obtain the 
HFL, we use the fact that
the entropy of Lovelock black holes is independent of matter content and given by \cite{Wald:1993nt, Iyer:1994ys}\footnote{{See \cite{Son:2013eea} for what happens with the HFL if one instead identifies $S$ with the black hole area, $S=A/4$.}}
\be
S=\frac{\Sigma_{d-2}}{4}\sum_{k=1}^K\alpha_k\frac{(d-2)!}{(d-2k-1)!}\frac{k\sigma^{k-1}}{d-2k}r_+^{d-2k}\,.
\ee
Upon multiplying both sides of the equation of state \eqref{EState} by
$
\delta V=\Sigma_{d-2}r_+^{d-2}\delta r_+\,,
$
and using that
\be
\delta S=\frac{\Sigma_{d-2}}{4}\sum_{k=1}^K\alpha_k\frac{ k\sigma^{k-1}(d-2)!}{(d-2k-1)!} r_+^{d-2k-1}\delta r_+\,,
\ee
the equation of state can be re-written as the HFL for Lovelock black holes
\be\label{first1}
\delta E=T\delta S-P\delta V\,,
\ee
where \be\label{Ene}
E=\frac{\Sigma_{d-2}}{16\pi}\sum_{k=1}^K\alpha_k\frac{\sigma^k(d-2)!}{(d-2k-1)!}r_+^{d-2k-1}
\ee
is regarded as an energy associated with the black hole, whose interpretation we discuss below.
This first law is equivalent to the equation of motion \eqref{Trr}  evaluated on the horizon.

The HFL can be `integrated' to give the following {\em Horizon Smarr Formula} (HSF):
\ba\label{ESmarr}
(d-3)E=(d-2)TS-(d-1)PV+\sum_{k=2}^K {2(k\!-\!1)}\alpha_k\Psi^{(k)},\nonumber\\
\Psi^{(k)}=\frac{\Sigma_{d-2}}{16\pi}\frac{\sigma^{k-1}(d-2)!}{(d-2k-1)!}r_+^{d-2k}\Bigl(\frac{\kappa}{r_+}-\frac{4\pi kT}{d-2k}\Bigr)\,, \ \qquad
\ea
where the `potentials' $\Psi^{(k)}$ are the thermodynamic conjugates to the $\alpha_k$ quantities.   Their presence (relevant for $K>1$) is required for  \eqref{ESmarr} to hold, and can be derived from the HFL by an Euler scaling argument, where $E=E(S,V)$ is regarded as a function of entropy and volume, provided the former is extended to include the variations of the couplings \cite{Kastor:2010gq} (that will not play a significant role in what follows).  This yields the term $(d-1)PV$
in \eqref{ESmarr}, since $ V\propto L^{d-1}$.

Having identified the horizon internal energy,
we can now define the {\em horizon enthalpy} $H$, and the {\em horizon Gibbs free energy} $G$ according to standard thermodynamic prescription,
\be\label{Gibbs}
H=E+PV\,,\quad G=E-TS+PV
\ee
and these  satisfy
\be\label{first2}
\delta H=T\delta S+V\delta P\,,\quad \delta G=-S\delta T+V\delta P
\ee
using the HFL  \eqref{first1}.
Note that insertion of \eqref{Gibbs} into \eqref{ESmarr}  yields
\be\label{HSmarr}
(d-3)H=(d-2)TS- 2PV+\sum_{k=2}^K {2(k-1)}\alpha_k\Psi^{(k)}\,.
\ee
Similar to the HFL, both the HSF \eqref{ESmarr} and its Legendre equivalent
\eqref{HSmarr} are valid irrespective of the matter content.  We note that the coefficient of the $PV$ term in
\eqref{HSmarr} has the same value as that in the extended phase space approach \cite{KastorEtal:2009, Altamirano:2014tva}.

Criticality and possible phase transitions depend on the behaviour of
\be
G=G(P,T)
\ee
which can be (parametrically) obtained by inverting  the equation of state, yielding
\ba
T&=&T(r_+,P)=\frac{4r_+}{K_\sigma}\bigl(P+P_\sigma)\,,\nonumber\\
G&=&G(r_+,P)=\frac{\Sigma_{d-2}}{d-1}Pr_+^{d-1}+\Sigma_{d-2}\sum_{k=1}^K\frac{\alpha_k(d-2)!}{(d-2k-1)!}\times\nonumber\\
&&\times \,r_+^{d-2k+1}\sigma^{k-1}\Bigl(\frac{\sigma}{16\pi r_+^2}-\frac{k}{d-2k}\frac{P+P_\sigma}{K_\sigma }\Bigr)\,,
\ea
where
\ba
P_\sigma&\equiv&\sum_{k=1}^K\frac{\alpha_k}{16\pi} {\frac{(d-2)!(d-2k-1)}{(d-2k-1)!}}\Bigl(\frac{\sigma}{r_+^2}\Bigr)^k\,,\nonumber\\
K_\sigma &\equiv&\sum_{k=1}^K\frac{k \alpha_k (d-2)!}{(d-2k-1)!}\Bigl(\frac{\sigma}{r_+^2}\Bigr)^{k-1}\,.
\ea

In this way one can study the behaviour of the Gibbs free energy and the potential criticality regardless of the actual knowledge of the
matter content of the theory.
We stress that $P$ is not necessarily positive (for example in the vacuum dS case $P$ has to be negative) and to map all the possible scenarios it makes sense to study all three cases of positive, zero, or negative pressure. 
 It is the actual matter content of a given theory that imposes associated restrictions on the
possible pressure interval  and gives the phase diagram a concrete physical interpretation, as we shall demonstrate in the sequel.

\section{$P-V$ criticality: some examples}

Before proceeding to a general comparison between horizon thermodynamics and the extended phase space approach, we shall consider some examples. Specifically, we  illustrate the possible behaviour of the horizon Gibbs free energy and the associated
variety of interesting phase transitions that occur in the horizon thermodynamics of spherically symmetric black holes in first few lower-order Lovelock gravities (small values of $K$), generalizing recent results for the Gauss-Bonnet case \cite{Ma:2015llh}.

\subsection{Einstein gravity}

We start with an example from  Einstein gravity $(K=1$) in $d=4$ dimensions (similar results hold in higher $d$).
Irrespective of the matter content, 
the equation of state \eqref{EState} reads
\be\label{EoSEin}
P=\frac{T}{2r_+}-\frac{\sigma}{8\pi r_+^2}\,,\quad V=\frac{\Sigma_2 r_+^3}{3}\,,
\ee
while the other thermodynamic quantities take the following explicit form:
\be
S=\frac{\Sigma_{2}r_+^2}{4}\,, \ E=\frac{\Sigma_2 \sigma r_+}{8\pi}\,, \ G=\frac{\Sigma_2 r_+}{6}\Bigl(\frac{3\sigma}{8\pi}-r_+^2 P\Bigr),
\ee
and satisfy the horizon first laws \eqref{first1} and \eqref{first2}.

\begin{figure}
\begin{center}
\includegraphics[width=0.47\textwidth,height=0.31\textheight]{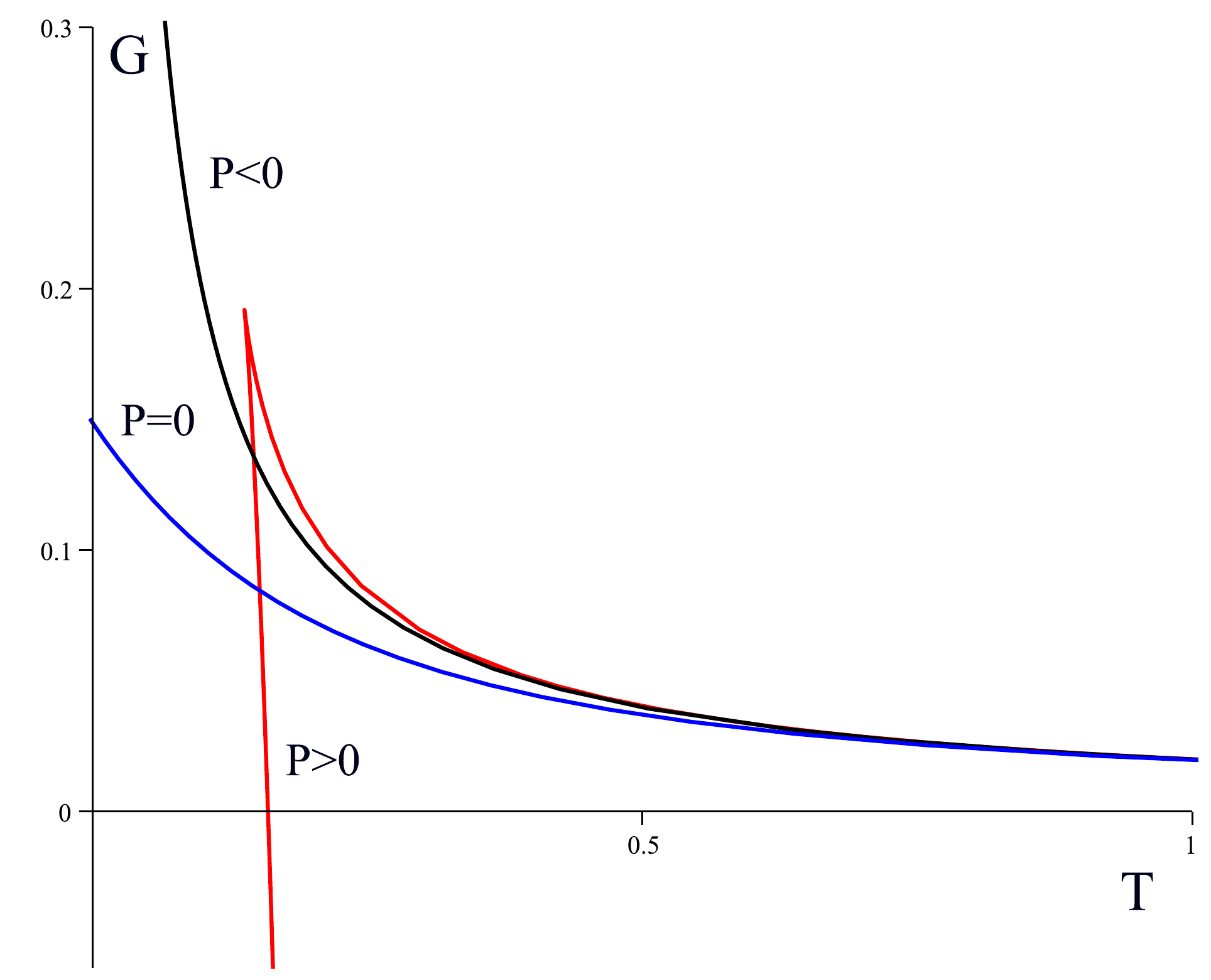}
\caption{\textbf{Horizon thermodynamics: $d=4$ spherical Einstein black holes.}
The $G-T$ diagram is displayed for $P=0.03$ (red curve), $P=0$ (black curve) and $P=-0.2$ (blue curve). For positive pressures we observe a characteristic shape reminiscent of the Hawking--Page behavior.
}
\label{Fig1}
\end{center}
\end{figure}
The behaviour of the horizon Gibbs free energy is for $\sigma=1$ displayed in Fig.~\ref{Fig1}. Whereas for $P>0$ we observe a shape characteristic
for the Hawking--Page transition of Schwarzschild-AdS black holes  \cite{Hawking:1982dh}
(illustrated in Fig.~\ref{Fig:HP}), for $P=0$ and $P<0$ we see that $G$ is relatively simple and respectively reminiscent of what happens for asymptotically dS and  asymptotically flat (uncharged) black holes \cite{Altamirano:2014tva, Kubiznak:2015bya}. However, this similarity is only superficial and the actual physical interpretation depends on the matter content of the theory, as we shall demonstrate below.  No other interesting phase behaviour is possible for $\sigma=1$.

\subsection{Gauss--Bonnet gravity}

\begin{figure}
\begin{center}
\includegraphics[width=0.47\textwidth,height=0.31\textheight]{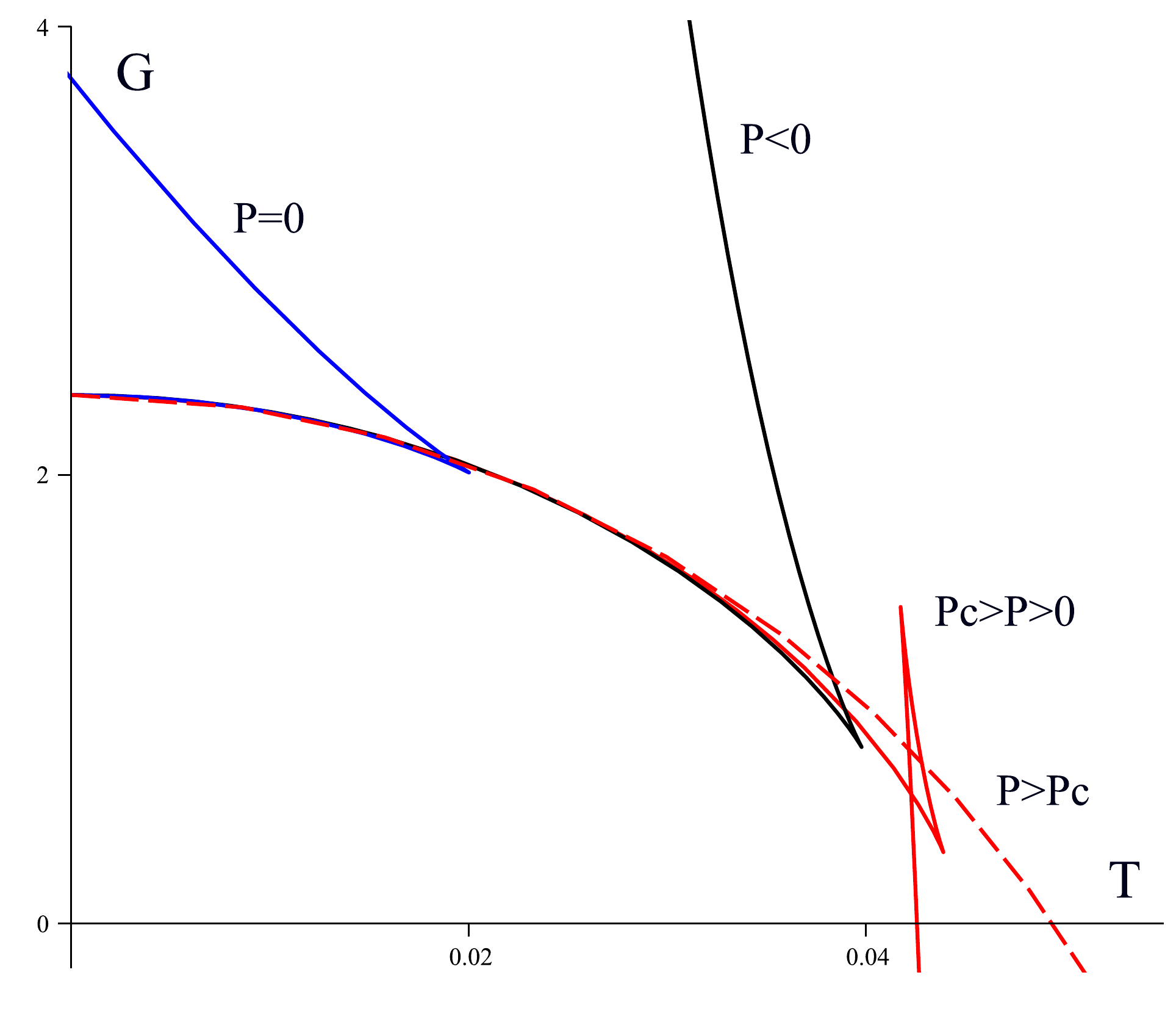}
\caption{\textbf{Horizon thermodynamics: $d=5$ spherical Gauss--Bonnet black holes.}
The $G-T$ diagram is displayed for $P=0.01$ (red dash curve), $P=0.0025$ (red solid curve), $P=0$ (black curve), and $P=-0.05$ (blue curve) and $\alpha_2=1$. For small positive pressures we observe a characteristic swallow tail  reminiscent of the Van der Waals-like phase transition.
}
\label{Fig:2}
\end{center}
\end{figure}

Carrying out the same analysis in Gauss--Bonnet gravity ($K=2$) in $d=5$ dimensions,  the equation of state reads
\be
	P = \frac{3T}{4r_+}-\frac{3\sigma}{8\pi r_+^2} + \frac{3\alpha_2\sigma T}{r_+^3}\,,\quad
V = \frac{\Sigma_3r_+^4}{4}\,,
\ee
while the other quantities are
\ba
S &=& \frac{\Sigma_3r_+^3}{4}\left(1\!+\!\frac{12\sigma\alpha_2}{r^2_+} \right)\,,
\quad
E = {\frac{3\Sigma_3\sigma r^2_+}{16\pi}\left(1\!+\!\frac{2\alpha_2\sigma}{r^2_+}\right)}\,,\nonumber\\
G&=& {
\frac{\Sigma_3[72\alpha_2^2\sigma\!-\!18\sigma r_+^2(\sigma\!+\!8\pi r_+^2 P)\alpha_2\!+\!3\sigma r_+^4\!-\!4\pi P r_+^6]}{48\pi(r_+^2+4\sigma \alpha_2)} },\nonumber\\
\ea
and satisfy the horizon first laws \eqref{first1} and \eqref{first2}.

The corresponding $G-T$ diagram for spherical ($\sigma=1$) black holes is displayed in Fig.~\ref{Fig:2}.
In contrast to the $K=1$ case, we now see that the additional gravitational non-linearity can yield more interesting phase behaviour.
Namely,
for sufficiently small positive pressures \cite{Ma:2015llh, Frassino:2014pha}
\be
0<P<P_c=\frac{1}{96 \pi \alpha_2}\,,
\ee
 we observe a characteristic swallow tail reminiscent of the Van der Waals-like phase transition for $d=4$ charged black holes in extended phase space
\cite{Kubiznak:2012wp}, illustrated in Fig.~\ref{Fig:VdW}.
For $P>P_c$ the swallow tail disappears and the Gibbs free energy becomes smooth. On the other hand for $P=0$ and $P<0$ we observe a cusp
(corresponding to a divergent specific heat) and the shape of $G=G(T)$ reminds that of the charged asymptotically dS and asymptotically flat black holes, c.f. \cite{Altamirano:2014tva, Kubiznak:2015bya}.


\subsection{Higher-order Lovelock gravity}

\begin{figure}
\begin{center}
\includegraphics[width=0.47\textwidth,height=0.31\textheight]{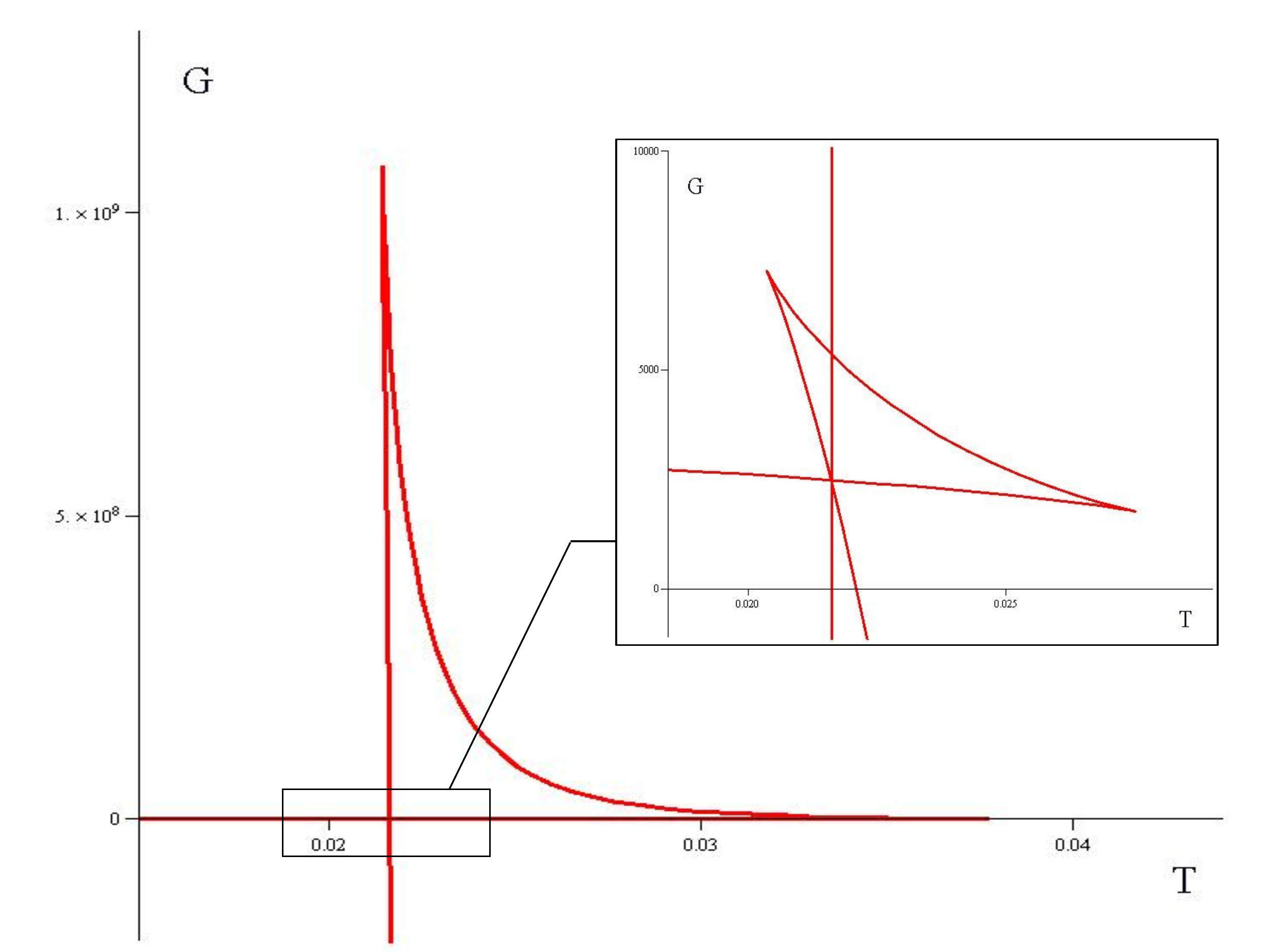}
\caption{{\textbf{Horizon thermodynamics: triple point.}
The $G-T$ diagram is displayed for a spherical black hole in the 4-th order Lovelock gravity for the following choice of parameters:
$\alpha_2=0.2, \alpha_3=2.8, \alpha_4=1, P=0.000425.$ We observe two swallow tails merging together, characterizing an existence of a triple point.}
}
\label{Fig:3}
\end{center}
\end{figure}

For $K > 2$ we find further interesting phase behaviour.   At each additional order in the Lovelock expansion, we gain an additional degree of freedom corresponding to the additional Lovelock coupling $\alpha_K$, allowing for more complex structures to arise.  We find phenomena similar to  those
 seen  previously in extended phase space thermodynamics for $K=1$, such as reentrant phase transitions
 \cite{Altamirano:2013ane},  double swallow tails and a corresponding triple point \cite{Altamirano:2013uqa},
 and even (for $K>2$)  isolated critical points \cite{Frassino:2014pha,Dolan:2014vba,Hennigar:2015esa}.
However in contrast to the extended phase space approach, such behaviour in horizon thermodynamics is entirely due to the non-linearity of gravity (the larger values of $K$), fully independent of the matter distribution. We depict a triple point in 4-th order Lovelock gravity in Fig.~\ref{Fig:3}.

 It remains an interesting open question whether the horizon thermodynamics of higher-order Lovelock theories can bring some additional qualitatively new phase transitions to those described in this section. In particular, can one find `$n$-tuple swallow tails' and the corresponding $n$-tuple critical points? We leave this question for future work.

\section{Comparison to extended thermodynamics with variable $\Lambda$}

In this section we shall compare  horizon thermodynamics to the recently studied  (canonical ensemble) extended
phase space thermodynamics of asymptotically AdS black holes.
The latter, sometimes referred to as black hole chemistry \cite{Kubiznak:2014zwa}, is essentially   `standard black hole thermodynamics' with the additional feature that
the (negative) cosmological constant is treated as an additional thermodynamic variable, which is interpreted
as a thermodynamic pressure $P_\Lambda$ according to Eq.~\eqref{PL} and allowed to vary in the
corresponding first law. The first law  for spherically symmetric Lovelock black holes then takes the following form \cite{Kastor:2010gq}:
\be\label{1st}
\delta M=T\delta S+\sum_i\Phi_i \delta Q_i+V_{\tiny\mbox{TD}}\delta P_\Lambda+\sum_{k=2}\Psi^{(k)}\delta \alpha_k\,,
\ee
and implies the associated Smarr formula
\ba\label{Smarr}
(d-3)M&=&(d-2)TS+(d-3)\sum_i\Phi_i Q_i-2V_{\tiny\mbox{TD}}P_\Lambda\nonumber\\
&& +\sum_{k=2}2(k-1)\Psi^{(k)}\alpha_k
\ea
through the Euler scaling argument. Here $M$ stands for the black hole mass, now interpreted as a {\em gravitational enthalpy}, distinct from the enthalpy defined in \eqref{Gibbs}.
We have also included the possibility that the black holes are multiply-charged  with several $U(1)$ charges $Q_i$  and
 corresponding electric potentials $\Phi_i$.
The horizon temperature $T$ and associated entropy $S$ are the same as in the horizon thermodynamics approach.

Let us now study some differences between the HFL \eqref{first1} and the extended first law \eqref{1st}.  The most obvious distinction is
the appearance of extra work terms, $\sum_i\Phi_i \delta Q_i$, in \eqref{1st}.\footnote{Note that in \eqref{1st} we have also included the possibility of variable $\delta\alpha_k$, which are needed to relate the first law to the Smarr relation \eqref{Smarr} through the Euler scaling argument. These terms are not natural in the horizon thermodynamics description. In what follows we simply ignore them even for the extended phase space description as they are not central for our further discussion.}  
These terms in the horizon case \eqref{first1} are instead interpreted as contributions to the pressure, which is  associated with all matter fields. In the extended case \eqref{1st} one only has a completely isotropic
pressure due to the cosmological constant.

A more important difference between \eqref{first1} and  \eqref{1st}  is the nature of the black hole volume.
In the horizon approach $V$ is assumed to be given by \eqref{Vol}; it is associated with the `Euclidean geometric volume' of the black hole and is independent of the matter content, c.f. \cite{Parikh:2005qs, Cvetic:2010jb, Ballik:2010rx, Ballik:2013uia}. In contrast to this the volume in extended thermodynamics
\be
V_{\tiny\mbox{TD}}=\Bigl(\frac{\partial M}{\partial P_\Lambda}\Bigr)_{S,Q_1,\dots }
\ee
is a {\em thermodynamic volume} \cite{Cvetic:2010jb}, a quantity conjugate to the pressure $P_\Lambda$.
Hence $V_{\tiny\mbox{TD}}$ is not an independent input but directly follows from the identification of the black hole mass. It can also depend on the matter content of the theory;  for example  the thermodynamic volumes of supergravity black holes have this feature \cite{Cvetic:2010jb}.

 Another important difference is the nature and distinction between  the quantities $E$, $H$, and $M$. Whereas the latter is the black hole mass and can be calculated by standard methods, e.g. the method of conformal completion \cite{Ashtekar2000,Das:2000cu}, the physical meaning of  $E$ is not clear.
It evidently plays the role of energy in \eqref{first1}, but
this quantity is not the mass of black hole; indeed its properties are quite different.  It vanishes for planar/toroidal black holes (for which $\sigma=0$) and can be negative for higher-genus topological/hyperbolic black holes (for which $\sigma=-1$). It has been noted that it is associated with  the transverse geometry of the horizon  \cite{Paranjape:2006ca}.

Since it is a function only of the  horizon curvature $\sigma$ and the horizon radius $r_+$,  we propose that it is  the {\em horizon curvature energy}: the energy required to warp space time so that it embeds an horizon.  This definition is analogous to that of the spatial curvature density in cosmology, which depends only on the curvature of spatial slices at constant time in an FRW cosmology.  Likewise, the horizon enthalpy $H$ then can be interpreted as the energy required to both warp spacetime and displace its matter content so that a black hole can be created.

 In particular, using   \eqref{HSmarr} and \eqref{Smarr}, we find the following relation between $M$ and  $H$:
\be
M=H+\sum_{i}Q_i\Phi_i+\frac{2}{d-3}\bigl(VP-V_{\tiny\mbox{TD}} P_{\Lambda}\bigr)
\ee
valid for the charged AdS Lovelock black holes.
For singly charged Lovelock black holes,    $V=V_{\tiny\mbox{TD}}$ \cite{Kastor:2010gq, Frassino:2014pha}
yielding
\be\label{Mm}
M=H+Q\Phi+\frac{2}{d-3}VP_m\,.
\ee
as the relationship between mass and horizon enthalpy $H$.

If no matter apart from a cosmological constant is present   $P_m=0$.  $H$ and $M$ then represent the same quantities, and so
\be\label{HMeq}
H=M=E+P_\Lambda V
\ee
which is the sum of the energy $E$ needed for warping the spacetime to embed the black hole horizon plus the energy $P_\Lambda V$ needed to place the black hole into a cosmological environment (`to displace the vacuum energy').
Note that for {\em planar black holes} $E$ vanishes and the mass is entirely given by the $P_\Lambda V$ term.

We pause to comment that the quantity $E$ is related to  the generalized Misner--Sharp mass $m_{\tiny \mbox{MS}}=m_{\tiny \mbox{MS}}(r)$   \cite{Maeda:2007uu, Maeda:2011ii}
\be
m_{\tiny \mbox{MS}}(r_+)=P_\Lambda V+E =M
\ee
evaluated on the black hole horizon \cite{Cai:2008mh}.  The last equality follows from \eqref{HMeq} (which holds for $P_m=0$),  and so we see that the mass of a Schwarzschild  AdS black hole is the Misner--Sharp mass on the horizon.  Setting $P_m\neq 0$, it has been shown that $m_{\tiny \mbox{MS}}(r_+)$ satisfies the HFL \cite{ Hayward:1997jp,Cai:2008mh}.

Criticality and possible phase transitions in the framework of extended phase space are governed by the
associated Gibbs free energy
\be
G_\Lambda=M-TS\,,
\ee
in  comparison to the horizon Gibbs free energy $G$  \eqref{Gibbs}.

In particular, and obvious from the above discussion, in the
vacuum with negative cosmological constant case we have the same expressions
\be
G={G_\Lambda}\,,\quad P=P_\Lambda
\ee
for the Gibbs free energy and  equation of state.  Only in this case and  for {\em positive $P$} do
the two approaches yield the same kind of thermodynamic behaviour and phase transitions
 (Van der Waals behaviour,  reentrant transitions, triple points,  isolated critical points) in any Lovelock theory.
These phenomena will only take place for sufficiently large $K$ (sufficient gravitational non-linearity).

The two approaches differ  significantly once matter is introduced.  Generically they give rise to very distinct phase diagrams with completely different physical interpretations. The difference is rooted in the inherent degeneracy in  horizon thermodynamics:  it is described by only two parameters $T$ and $P$ (together with their conjugates).  This degeneracy is removed in  extended phase space thermodynamics, with each matter field having its own contribution to the free-energy,
leading to a description in a different (often incompatible) thermodynamic ensemble.  Furthermore, in horizon thermodynamics negative pressures are possible even if $\Lambda <0$, whereas in the extended case negative pressure requires $\Lambda > 0$.

We shall now illustrate these distinctions for a
spherical $(\sigma=1)$ charged-AdS black hole in $d=4$ dimensions $(K=1$)
\ba
ds^2&=&-fdt^2+\frac{dr^2}{f}+r^2d\Omega_2^2\,,\nonumber\\
F&=&dA\,,\quad A=-\frac{Q}{r}dt\,,
\ea
 where $d\Omega^2=r^2(d\theta^2+\sin^2\!\theta d\varphi^2)$,
\be\label{f}
f=1-\frac{2M}{r}+\frac{Q^2}{r^2}+\frac{r^2}{l^2}\,,
\ee
 and $\Lambda=-\frac{3}{l^2}\,$ is the cosmological constant.  This simple example will allow us to discuss all important differences without the need for complicated expressions;  generalization to `arbitrary' charged Lovelock black holes is straightforward   \cite{Frassino:2014pha}.

The  HES \eqref{EState} now reads
\be\label{V4d}
P=\frac{T}{2r_+}-\frac{1}{8\pi r_+^2}\,,\quad
V=\frac{4}{3}\pi r_+^3\,,
\ee
upon setting  $\sigma=1$ in \eqref{EoSEin}.
Interestingly, using the expression for the energy-momentum tensor,
\be\label{P4d}
P_m=T^r{}_r=-\frac{Q^2}{8\pi r_+^4}\,,
\ee
the HES \eqref{V4d}
 can be rewritten as
\be\label{RNEoS}
P_\Lambda=\frac{T}{2r_+}-\frac{1}{8\pi r_+^2}-P_m=
\frac{T}{2r_+}-\frac{1}{8\pi r_+^2}+\frac{Q^2}{8\pi r_+^4}\,,
\ee
which is the   extended phase space equation of state in the canonical ensemble \cite{Kubiznak:2012wp} upon  setting $Q$ constant and identifying $P_\Lambda=-\Lambda/(8\pi)$.   Note that  $V_\Lambda=V$ and so the thermodynamic  and geometric volumes are the same and
\be
P_\Lambda = P+\frac{Q^2}{8\pi r_+^4}
\ee
since $P=P_m+P_\Lambda$.

Note that in the extended phase space approach  there is no need to  `invoke the Einstein equations' to derive this equation of state since we are using a concrete solution.   In fact \eqref{RNEoS} simply follows from the `definition' of the temperature
\be
T=\frac{f'}{4\pi}\,,
\ee
upon using the explicit form of $f$ from \eqref{f}.
The horizon enthalpy
\ba
H&=&\frac{r_+(1+2\pi T r_+)}{3}
\ea
and mass (gravitational enthalpy)
\be
M=\frac{r_+^2l^2+Q^2l^2 +r_+^4}{2l^2 r_+}
\ee
of the black hole are related via
 \eqref{Mm}, $M=H+\Phi Q+2VP_m$\,, where $\Phi=Q/r_+$, and $P_m$ and $V$ are given by
\eqref{P4d} and \eqref{V4d}.
This then implies the following relation:
\ba
G_\Lambda&=&G+\Phi Q+2 VP_m=G+\frac{2}{3}\frac{Q^2}{r_+}\,,\nonumber\\
P_\Lambda&=&P+\frac{Q^2}{8\pi r_+^4}
\ea
between the horizon and extended Gibbs free energies.\footnote{ Note that the extended phase space equation of state \eqref{RNEoS} was directly derived from the horizon equation of state  \eqref{EoSEin} by splitting $P=P_m+P_\Lambda$,. This is not true for the Gibbs free energy ${G_\Lambda}$. }

\begin{figure}
\begin{center}
\includegraphics[width=0.47\textwidth,height=0.31\textheight]{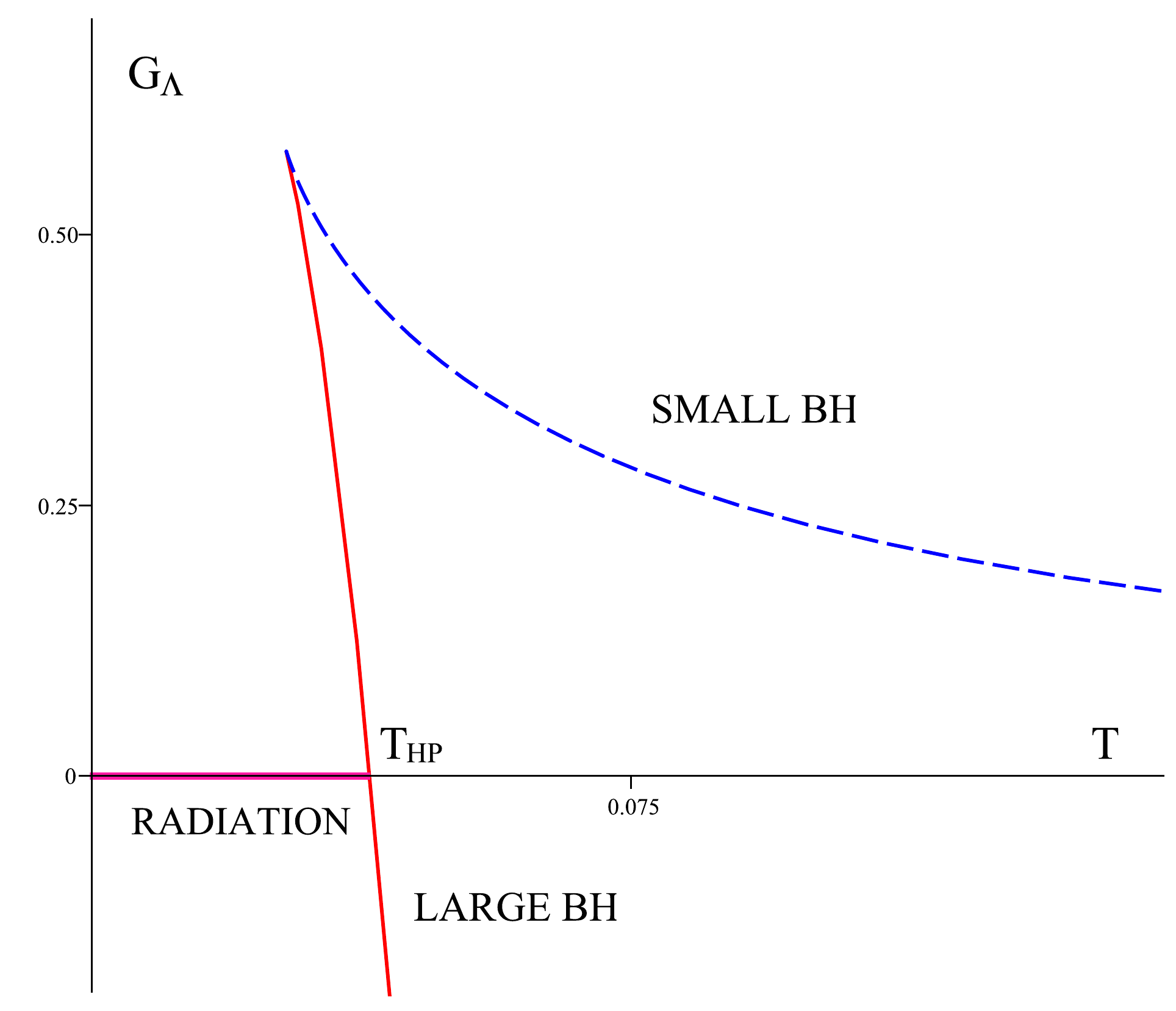}
\caption{ \textbf{Hawking--Page transition.}
The characteristic $G_\Lambda-T$ diagram is displayed for the uncharged ($Q=0$) AdS spherical black hole in $d=4$.
The black hole Gibbs free energy admits two branches of black holes: small black holes (displayed by blue dashed curve)
have negative specific heat and are thermodynamically unstable while large black holes (solid red curve) have positive specific heat and thermodynamically dominate for large temperatures, $T>T_{\tiny\mbox{HP}}$, over the radiation phase displayed by horizontal magenta line.
Note that (being in the framework of extended phase space thermodynamics) each point on the black hole curve corresponds to different black holes (of increasing horizon radius $r_+$ from right on the dashed blue curve to bottom left) in the same environment of fixed $\Lambda$ and fixed $Q=0$.
}
\label{Fig:HP}
\end{center}
\end{figure}

\begin{figure}
\begin{center}
\includegraphics[width=0.47\textwidth,height=0.31\textheight]{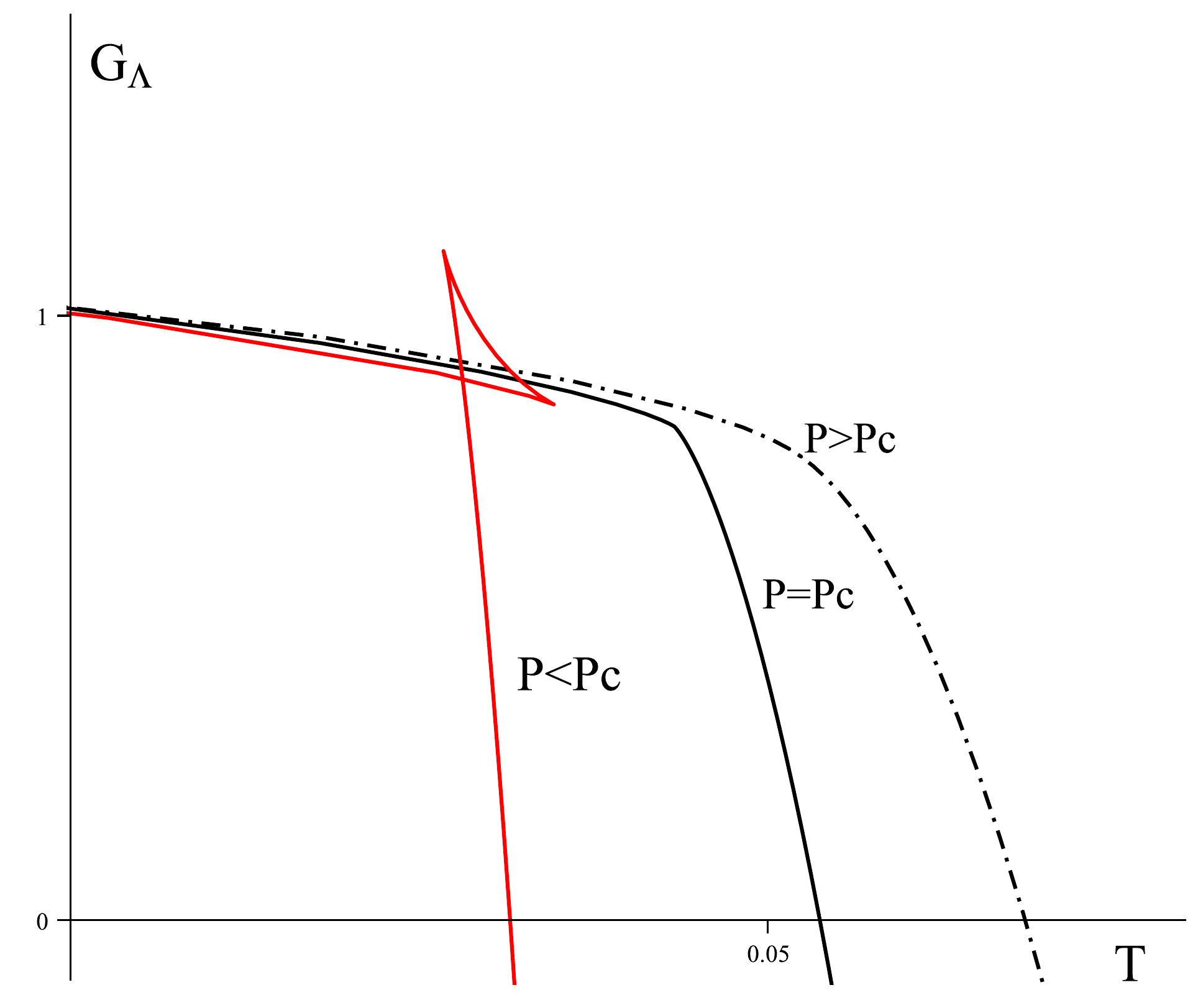}
\caption{\textbf{Van der Waals-like phase transition.}
The characteristic $G_\Lambda-T$ diagram is displayed for the charged ($Q=1$) AdS spherical black hole in $d=4$.
For sufficiently small pressures, $P<P_c=1/[96\pi Q^2]$, the $G_\Lambda-T$ diagram displays the characteristic swallow tail behaviour indicating a small to large black hole phase transition ala Van der Waals. As with Fig. \ref{Fig:HP} , each point on the curve corresponds to different black holes (of increasing horizon radius $r_+$ from left to bottom right) in the same environment of fixed $\Lambda$ and $Q$.
}
\label{Fig:VdW}
\end{center}
\end{figure}

These relations imply fundamentally different thermodynamic behaviour in the two approaches.
 Even after
removing  the degeneracy in  \eqref{V4d} by imposing a constant $Q$ constraint, the $P=const$ and $P_\Lambda=const$
slices of thermodynamic phase space  are incompatible, and yield different behaviour of the Gibbs free energies $G(T)$ and $G_\Lambda(T)$. We shall illustrate this point by comparing the positive pressure curve in Fig.~1 describing the behaviour of $G$ in horizon thermodynamics  to  that  of $G_\Lambda$ displaying the Hawking--Page transition for $Q=0$ and the Van der Waals like behavior for $Q\neq 0$ in the extended phase space thermodynamics, Fig.~\ref{Fig:HP} and Fig.~\ref{Fig:VdW}.

In horizon thermodynamics the description is in terms only of $\{T,P\}$, and only
`Hawking--Page-like behavior' of the horizon Gibbs free energy $G=G(P,T)$ can be observed, as shown in
Fig.~\ref{Fig1}.  Furthermore, as $T$ changes, moving along a constant-$P$ curve entails modifying some combination of $Q$, $r_+$, and $\Lambda$:  different points on the curve are comparing different black holes in {\em different environments}\footnote{Since constant-$P$ undetermined condition, its realization can be always achieved by setting $Q=0$ and tuning $\Lambda$ accordingly. For this reason it is not that surprising that the horizon Gibbs free energy mimics the $Q=0$ behavior of the extended phase space Gibbs free energy.}.  The expected transition at  $G=0$ to pure radiation (which has $Q=0$) can only occur if there is a reservoir of charge, so that $Q$ can appropriately vanish as this transition takes place.

In other words, the physical interpretation of Fig.~\ref{Fig1} in horizon thermodynamics depends crucially on the matter content.  In contrast to this, the extended phase-space picture breaks this degeneracy, allowing for imposition of independent constraints on $Q$ and the pressure $P_\Lambda$.  If  $Q=0$ (Fig.~\ref{Fig:HP}) the standard Hawking--Page phase transition is recovered \cite{Kubiznak:2014zwa}, whereas for fixed $Q\neq 0$ (Fig.~\ref{Fig:VdW}), Van der Waals-like behaviour is observed \cite{Kubiznak:2012wp}, with the  Gibbs free energy $G_\Lambda=G_\Lambda(P_\Lambda,T, Q)$ exhibiting a
swallowtail structure.  In either case, each point on the curve in a $G_\Lambda$ vs. $T$ diagram corresponds to different black holes in the {\it same environment} (the same $\Lambda$ and $Q$).

We see  that the distinction between the two approaches in this example is reminiscent of the canonical vs. grand-canonical description of charged AdS black holes. For a charged AdS black hole we observe  Van der Waals phase transitions only in a canonical (fixed $Q$) ensemble (as in the extended phase space approach), whereas in the grand canonical (fixed $\Phi$) ensemble behaviour similar to Fig.~\ref{Fig1} is observed (as in horizon thermodynamics).

In summary, horizon thermodynamics describes a system from the viewpoint of an ensemble described by only two variables $P$ and $T$.  The   Gibbs free energy therefore only depends on the type of gravity considered.  Such a description is `universal' and `formally independent' of the matter content. However, the actual interpretation of the thermodynamic behaviour is matter dependent. In general it is not unique  due to the degeneracy of the description, in contrast to the non-degenerate description in extended phase space thermodynamics. Even after the degeneracy is removed,   horizon thermodynamics often leads to a different ensemble description, incompatible with  extended phase space thermodynamics.

\section{Discussion}

We have reviewed the horizon thermodynamics approach to the thermodynamics of spherically symmetric black holes in Lovelock gravity and compared  it to the extended phase space approach.
The key idea of horizon thermodynamics  is to rewrite the Einstein equations evaluated on the black hole horizon as a thermodynamic identity, obtaining an
horizon equation of state together with a first law of horizon thermodynamics. The explicit form of this law depends on identifying the black hole volume, black hole entropy, and the temperature. The first law then defines a quantity $E$, which we have proposed is the {\em horizon curvature energy}: the energy required to warp space time so that it embeds an horizon.
By employing an Euler scaling argument we also derived the Horizon Smarr Formula \eqref{ESmarr}, as well as obtained the horizon enthalpy $H$ and Gibbs free energy $G$. The latter allows one to study $P-V$ criticality in horizon thermodynamics.

Comparing this to the recently studied $P-V$ criticality in the context of asymptotically AdS black holes (so-called black hole chemistry \cite{Kubiznak:2014zwa}),  we find that the two approaches are quite different, in general leading to incompatible  thermodynamic descriptions of the same system.  Horizon thermodynamics
intrinsically contains a degeneracy amongst thermodynamic variables that are distinct in the extended phase space approach.  Only in the vacuum with negative cosmological constant do the two approaches lead to identical thermodynamics.

We have also shown that increasing non-linearity in the gravitational sector yields more interesting thermodynamic behaviour, and in this sense it is possible  in horizon thermodynamics to recover phenomena previously observed in black hole chemistry.  While this description might appear to be `universal' and `formally independent' of the matter content, in fact the interpretation of these phenomena  in horizon thermodynamics  will depend on the matter content of the theory.

Our study opens the possibility for studying $P-V$ criticality and associated phase transitions of black holes in various theories in the horizon thermodynamics context. Whereas in this paper we have concentrated on spherically symmetric black holes in Lovelock gravity, an interesting future study would be to consider a similar investigations for black holes in Lifshitz, $f(R)$, quasi-topological, and other theories of gravity. Another interesting future direction would be to go beyond the realm of black hole thermodynamics and consider for example the criticality of horizon thermodynamics for acceleration and cosmological horizons. If horizon thermodynamics indeed elicits universal features of `any horizon',   $P-V$ criticality should be a universal feature of all gravitational theories.


\section*{Acknowledgements}
This research was supported in part by Perimeter Institute for Theoretical Physics and by the Natural Sciences and Engineering Research Council of Canada. Research at Perimeter Institute is supported by the Government of Canada through Industry Canada and by the Province of Ontario through the Ministry of Research and Innovation.


\providecommand{\href}[2]{#2}\begingroup\raggedright\endgroup

\end{document}